\documentclass[reprint,superscriptaddress,amsmath,amssymb,aps,prd,twocolumn,floatfix,nofootinbib]{revtex4-2}
\usepackage{amsmath, amssymb}
\usepackage{listings}
\usepackage{graphicx}
\usepackage{dcolumn}
\usepackage{bm}
\usepackage{float,color}
\usepackage{xcolor}
\usepackage[normalem]{ulem}
\usepackage{tabularx}
\usepackage{mathtools} 
\usepackage{multirow}
\usepackage{amssymb}
\usepackage[caption=false]{subfig}
\usepackage{placeins}
\usepackage{scrextend}

\usepackage[colorlinks=true,linkcolor=blue,citecolor=blue,urlcolor=blue]{hyperref}

\begin{document}
	
\title{
Forward Bayesian Inference for the Binary Black Hole Populations
}	
\author{Poulami Dutta Roy}
\email{duttaroy.poulami@ufl.edu}
\affiliation{Department of Physics, University of Florida, PO Box 118440, Gainesville, Florida 32611-8440, USA}

\author{Tanmaya Mishra}
\affiliation{Institute of Cosmology and Gravitation, University of Portsmouth, Portsmouth, PO1 3FX, UK}
\affiliation{Department of Physics, University of Florida, PO Box 118440, Gainesville, Florida 32611-8440, USA}

\author{Achal Kumar}
\affiliation{Department of Physics, University of Florida, PO Box 118440, Gainesville, Florida 32611-8440, USA}

\author{Sergey Klimenko}
\affiliation{Department of Physics, University of Florida, PO Box 118440, Gainesville, Florida 32611-8440, USA}

\begin{abstract}
We present \texttt{PopCWB}, a Bayesian framework for the inference of the binary black hole (BBH) population based on the unmodeled search pipeline coherent WaveBurst (cWB). The standard population analysis takes an inverse approach by inferring the underlying population model parameters  for a given set of detected events using the individual posterior samples. In \texttt{PopCWB}, we take a forward approach to population inference by varying population models to identify the most optimal one which describes the observed events. Rather than using event-level posterior samples, \texttt{PopCWB} uses a large set of simulated events reconstructed by cWB and their total mass, estimated with a machine-learning regression, to construct marginal probability distributions. These marginal probabilities are then used in the construction of the likelihood.
We apply \texttt{PopCWB} to BBH events detected by cWB during the third observing run (O3) of the LIGO--Virgo--KAGRA detectors. We constrain an astrophysically motivated BBH population model that incorporates the effects of pulsational pair-instability supernovae and dynamical mergers. The analysis  predicts a maximum black hole mass of $\mathbf{48.3^{+8.2}_{-10.2} M_{\odot}}$ from stellar evolution and an inferred local BBH merger rate of $\mathbf{29.1^{+32.8}_{-16.0} \, Gpc^{-3} yr^{-1}}$. We compare these results with the existing population models from the literature.

\end{abstract}

\maketitle
	
\section{Introduction}
\label{sec:intro}

The detection of gravitational waves (GWs) from merging compact binaries by
the Advanced LIGO-Virgo-KAGRA (LVK) detectors \cite{KAGRA:2013rdx, Capote:2024rmo, LIGO:2024kkz, VIRGO:2026okv, KAGRA:2020tym} has established a new means of studying the
population of stellar-mass black holes. With the ensemble of detections, the majority of which are binary black holes
(BBHs) \cite{LIGOScientific:2026wfs}, we can move on from studying properties of individual binaries to probing the astrophysics of stellar evolution, the channels
through which BBHs form and merge, and the rate at which they do so across
cosmic time~\cite{LIGOScientific:2020kqk,KAGRA:2021duu, LIGOScientific:2026ctl}. 
A central question in population analysis is the mass distribution followed by black holes in the universe. This distribution is now known to depart from a simple power law describing the distribution of progenitor stars.
The inferred primary black hole mass spectrum shows an
excess near $\sim\!35\,M_\odot$~\cite{LIGOScientific:2020kqk,KAGRA:2021duu,Talbot:2018cva,Edelman:2021zkw}, whose
physical origin remains to be understood. Some studies associate the excess in the mass spectrum to pulsational pair-instability (PPI) supernovae experienced by a progenitor during the course of stellar evolution~\cite{Talbot:2018cva,Stevenson:2019rcw}. Unlike the pair-instability supernovae (PISN), where a massive star \cite{Fowler:1964,Rakavy:1967} experiences a violent explosion leaving no compact remnant, the PPI triggers a series nuclear pulses making the star lose mass but not strong enough to disrupt it \cite{Heger:2001cd, Woosley:2016hmi,Farmer:2020xne,Woosley:2019,Woosley:2021xba}. Due to these pulsations, stars of a wide range of mass, produce black holes within a narrow black hole mass region leading to the excess over the power law spectrum. The same mechanism sets the
maximum black hole mass attainable from stellar evolution, $\rm M_{BH}^{max}$,
marking the lower edge of the black hole mass gap \cite{Tong:2025wpz,Antonini:2025ilj,Ray:2025xti,Mould:2026sww,Karathanasis:2022rtr}. An accurate estimation of $\rm M_{BH}^{max}$ is crucial as it provides constraint on the $\rm ^{12}C(\alpha, \gamma) ^{16}O$ reaction rates occurring in the stellar core. The rate of this reaction determines the carbon/oxygen fraction in the star's core which in turn controls the probability of the star undergoing PPI. A higher rate creates more suitable circumstances for the PPI phenomenon to dominate the star's evolution, hence controlling the $\rm M_{BH}^{max}$. Stellar evolution simulations have predicted the $\rm M_{BH}^{max} \sim 45 M_{\odot}$ which remains robust against variations in factors like metallicity, stellar wind mass loss but susceptible to uncertainties in the $\rm ^{12}C(\alpha, \gamma) ^{16}O$ rate \cite{Farmer:2019jed, Woosley:2021xba, Takahashi:2018kkb, Baxter:2021swn,Mehta:2021fgz}. 
This highlights the importance of an accurate inference of the black hole mass spectrum which will provide meaningful constraints on the nuclear astrophysics.

Another aspect of the black hole mass spectrum considered in this study is the contribution from the other formation channels. The pair-instability mass gap can be populated by the hierarchical mergers which produce higher-generation black holes
\cite{Miller:2002pg,Gerosa:2021mno,Kimball:2020qyd,Plunkett:2026pxt,Afroz:2024fzp}. The mass spectrum can include the dynamically formed black holes where binaries are assembled through gravitational interactions in dense environments like active galactic nuclei disks \cite{Bartos:2016dgn,Mckernan:2017ssq, Rodriguez:2016avt,Antonini:2016gqe,Park:2017zgj,Samsing:2020qqd}.
Also, along with dynamically formed black holes, very massive stars are expected to undergo direct collapse to form black holes \cite{Bond:1984,Sakurai:2017opi}. 
In this study, we propose astrophysically motivated mass distribution that takes into account the contribution of black holes formed both from stellar-evolution and dynamical mergers.

Performing a population analysis on the large number of BBH detections helps infer the black hole mass spectrum. In literature, the standard Bayesian population inference takes an inverse approach by using the posterior samples obtained from parameter estimation of individual events to define the likelihood and infer the underlying population model \cite{Thrane_2019,Vitale:2020aaz,Tiwari:2017ndi,Mandel:2018mve,Farr:2019rap}. In this study, we introduce a {\em forward} Bayesian inference framework, called \texttt{PopCWB}, based on the unmodeled search pipeline, coherent WaveBurst (cWB)~\cite{Klimenko:2008fu,Klimenko:2015ypf, Klimenko:2022nji}. \texttt{PopCWB} varies population models and using a large simulation set it infers the most optimal model parameters that best describes the observed events. The simulation set consists of BBH signals, reconstructed by cWB, accommodating a wide range of black hole systems. Without relying on individual event posterior samples, \texttt{PopCWB} uses the total mass estimated with a machine-learning regression model and constructs the marginal probabilities. The marginal probabilities along with the information of the real events defines the Bayesian likelihood used in \texttt{PopCWB} to constrain the black hole mass model parameters.

The use of the unmodeled approach to detect and reconstruct signals in \texttt{PopCWB} removes the possibility of waveform systematics affecting the analysis. Additionally, the observed events and the simulations that define the likelihood are processed by the same analysis pipeline, 
so that any bias introduced by the search or the mass estimator acts on all triggers in the same way and largely cancels in the inference. Another approach of circumventing per-event parameter estimation by directly performing population inference from strain data has been introduced in \cite{Leyde:2026hvm}.

We apply \texttt{PopCWB} to analyze two
mass models: the Power-Law$+$Peak (PLP) model \cite{KAGRA:2021duu} adopted by the LVK
and the new cWB-PPI model, which accounts for the PPI phenomenon. Within the \texttt{PopCWB} analysis framework, the two mass models are compared through a Bayes factor estimation along with inferring astrophysically relevant parameters controlling the PPI and PISN physics.
The remainder of the paper is organized as follows. 
Section~\ref{sec:methodology} details the \texttt{PopCWB}
methodology, the cWB reconstruction and XGBoost mass estimation, the injection
set, and both the population models. Section~\ref{sec:results} presents the
inferred hyper-parameters, the framework cross-check against the LVK PLP result,
the model comparison, and the local merger rates followed by discussions and conclusions in
Section~\ref{sec:conclusions}.

\section{Methodology} \label{sec:methodology}

\subsection{Coherent WaveBurst} \label{sec:cwb}
The unmodeled search pipeline cWB is used in this analysis to identify and reconstruct GW signals observed in the Hanford-Livingston detector network \cite{Klimenko:2008fu,Klimenko:2015ypf}. The version of cWB used in our study searches for coherent power in time-frequency (TF) domain using multi-resolution WaveScan transform \cite{Klimenko:2022nji}. The algorithm then identifies excess power TF pixels above the detector noise and clustering them together. A coincident cluster across the detector network is identified as a trigger, associated with certain summary statistics describing the features of that trigger. For example, the summary statistics carry information about  the TF evolution, signal strength and coherence along with estimates on number of cycles contributing in the reconstructed waveform with further details given in \cite{Mishra:2024zzs}. 

In \texttt{PopCWB}, for each event, we perform parameter estimation using the cWB summary statistics and a supervised machine-learning algorithm XGBoost \cite{Chen_2016,OBrien:2021sua}. The XGBoost model is trained on the cWB summary statistics of a simulated set of triggers injected in O3 detector noise. This training set forms a small subset of the large sample of simulations used in the analysis (see Sec.\ref{sec:injections}). 
A detailed explanation of the training and testing of the XGB regression model can be found in Appendix \ref{sec:xgb_training}. For each event, the XGBoost model provides a point-estimate of the associated source-frame total mass which plays a crucial role in defining the likelihood of this analysis.

The 36 events observed in the O3 run of the LVK detectors, i.e. events reported in the GWTC-2.1 \cite{LIGOScientific:2021usb} and GWTC-3 \cite{KAGRA:2021vkt} catalog and reported with a significance of IFAR $> 1$ yr by the cWB pipeline, are included in the analysis.

\subsection{Simulations}\label{sec:injections}

\begin{figure}[t]
    \includegraphics[scale=0.35]{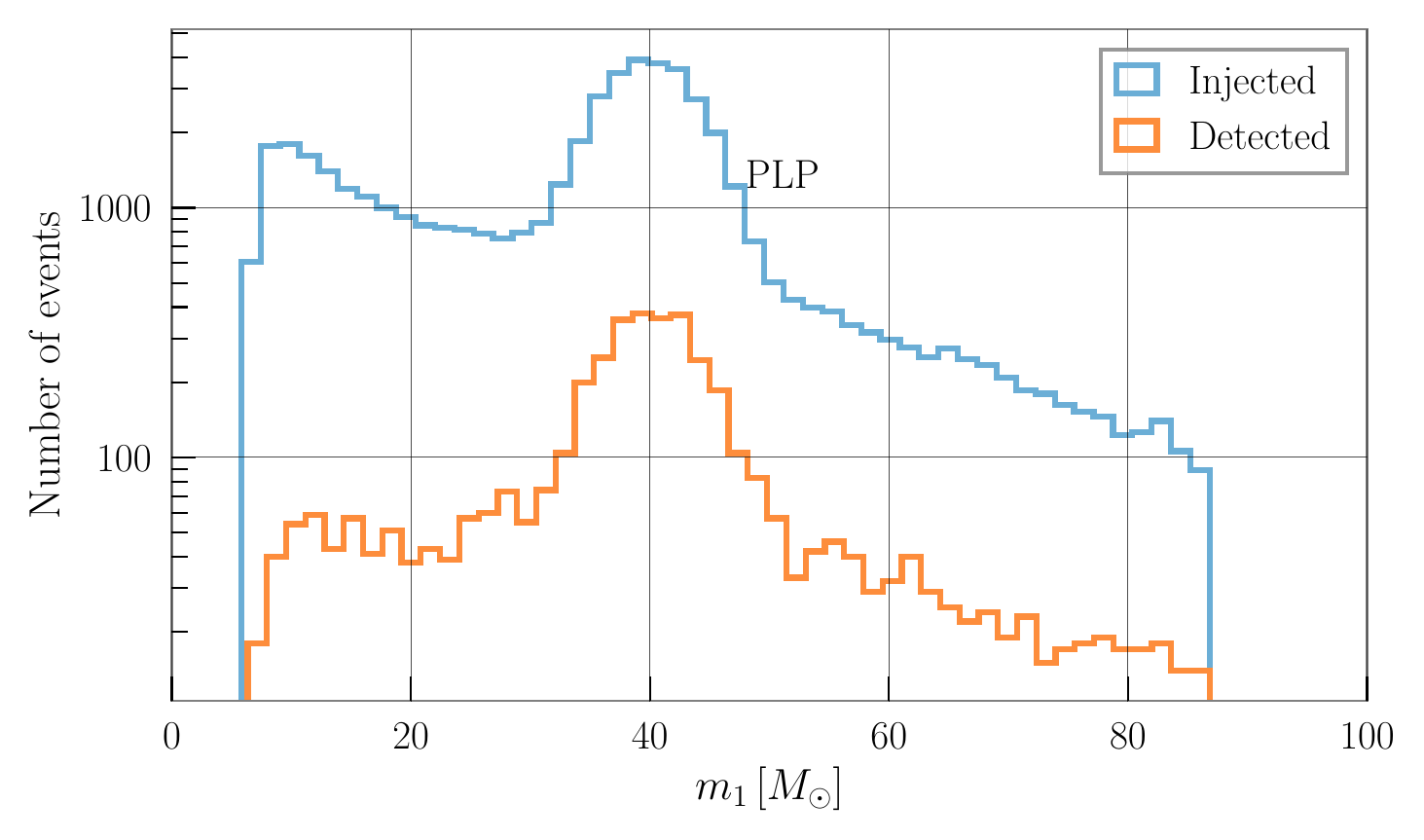}
    \caption{
    The distribution of primary mass of the reweighted injections following the PLP population model with the parameters having the median values as reported in GWTC3 analysis \cite{KAGRA:2021duu}. The blue curve indicates modeled distribution of injections with SNR$>6$, while the orange shows injections detected by cWB with $\rm IFAR > 1 yr$.
    }
    \label{fig:pop_dist}
\end{figure}

The simulated BBH signals are injected into the detector strain data and recovered with the cWB pipeline. For this analysis, we utilize a standard set of BBH injections which have been used by the LVK for the estimation of the sensitive volume in GWTC-3 \cite{KAGRA:2021duu,zenodo}. This set spans a component mass range of $1-100 \rm M_{\odot}$ with redshifts up to $\rm z =1.9$ \cite{zenodo}. The spins of the injections are isotropic in orientation and uniform in magnitude. The random draws from the injection set are reweighted \cite{Tiwari:2017ndi,Farr:2019rap,Essick:2021} to match a chosen population model. To speed up the computations, the injections with the network SNR less than 6 are excluded from the analysis.  For example, Fig.\ref{fig:pop_dist}, shows the injected and detected distributions of the primary component mass for the Power Law $+$ Peak model (see Sec.\ref{sec:PLP}).  The detected distribution, obtained by applying the cWB detection threshold of $\rm IFAR > 1 yr$, is used for the population inference. It accounts for the cWB selection effects, which vary across the mass range considered for the analysis. 

In \texttt{PopCWB}, the injections play two important roles. (i) The XGBoost regression model used for estimation of $\rm M_{cwb}$ is trained on the summary statistics and total injected mass of a subset of these injections. (ii) The likelihood defined in \texttt{PopCWB} is a product of marginal probabilities calculated from triggers drawn from the injection set as explained in Sec.\ref{sec:likelihood}.

\subsection{The Bayesian Approach}\label{sec:likelihood}
One of the main contributions of this work is the introduction of a new Bayesian framework for inferring the parameters of BBH population models. The key distinction from the standard approach, implemented by LVK, lies in the use of simulated injections and the construction of the likelihood.
For a given set of population parameters, $\Lambda$, the analysis draws BBH events from the simulation by assigning importance weights based on the component masses and redshift. Each draw contains $N_d=36$ events corresponding to the catalog of the O3 events observed by cWB with IFAR$> 1~\rm yr$. 

Similar to real detections, the simulated events in each draw are ranked in descending order according to the value of their total mass $\rm M_{cwb}$.
The above procedure is repeated $N_p$ times yielding a distribution of the total mass for each rank. 
A kernel density estimate (KDE) of this distribution defines the marginal probability $\rm p_i( M^i_{cwb} | \Lambda)$  calculated for the $\rm {M^i_{cwb}}$ of the corresponding detected event with the rank $\rm i$. The value of $N_p = 8000$ is selected to ensure that the KDE sampling errors are negligible. The product of marginal probabilities evaluated for the parameters $\Lambda$ defines the likelihood in the \texttt{PopCWB} analysis
\begin{equation} \label{eq:likelihood}
    \mathcal{L}(\Lambda) = \rm \Pi_{i=1}^{N_d} \, p_i ( M^i_{cwb} | \Lambda) . 
\end{equation}

In order to find the optimal set of population parameters describing the data, \texttt{BILBY} \cite{Ashton:2018jfp,Romero-Shaw:2020owr,Smith:2019ucc} is used with the nested sampling algorithm \texttt{DYNESTY} \cite{Speagle:2019ivv} to sample over the likelihood in eq.(\ref{eq:likelihood}) using flat priors.

The \texttt{PopCWB} analysis can be performed with the total mass of the binary, $\rm M_{cwb}$, being either in the source-frame or the detector-frame. Both quantities can be estimated using an accordingly trained regression model. In this work, we use the source-frame mass, making the inference insensitive to the BH redshift distribution. The redshift distribution is kept fixed (see next section) while performing inference on the BH population mass model. In this case, the choice of redshift distribution does not affect the mass model and only enters the calculation of the merger rates. In contrast, an analysis based on the detector-frame mass depends on the choice of redshift distribution, and the inference then requires simultaneous estimation of the mass model and the redshift distribution parameters. 

The proposed Bayesian framework has two additional advantages.
(i) Unlike the standard population analyses, \texttt{PopCWB} does not rely on the assumption of Gaussian detector noise at any stage of the inference. Instead, detector noise and signal reconstruction uncertainties are incorporated through the empirical distributions of the reconstructed black hole masses.
(ii) \texttt{PopCWB} accounts for selection effects directly by applying the same detection and reconstruction pipeline to both simulated injections and observed GW events.


The \texttt{PopCWB} uses a composite likelihood constructed from the marginal densities of the ranked events. The rank order procedure introduces correlations between neighboring ranks, which can lead to an underestimation of the parameter uncertainties. To account for correlations, we calibrate the posterior distributions empirically, 
following the procedure of~\cite{Kuleshov:2018, Rodrigues:2018, Talts:2018}. We generate $K=500$ mock catalogs, each containing $N_d$ events randomly selected from simulations of a population model with parameters $\Lambda_k$. For each catalog, $\Lambda_k$ is drawn jointly from the posterior samples obtained from the population inference using O3 events. 
For all mock catalogs, we apply the same population inference procedure as for the observed O3 catalog.
 To avoid evaluating the KDEs on events used to generate the mock catalogs, we randomly split the injection set into two equal subsets: one is used to generate the mock catalogs and the other to construct the likelihood. From the resulting 500 posterior sample sets, we compute, for each model parameter $\theta$ and catalog $k$, the fraction  $F_k$, of posterior samples below the true value $\theta_k$. For a calibrated posterior the $F_k$ are uniformly distributed on $[0,1]$. Departures from uniformity quantify the miscalibration and are encoded in the empirical calibration map
\begin{equation}
    C(f) = \frac{1}{K}\sum_{k=1}^{K} \Theta\left(f - F_k\right),
    \label{eq:calibration_map}
\end{equation}
where $\Theta$ is the Heaviside step function. If $F(\theta)$ denotes the posterior cumulative distribution of $\theta$ obtained with the observed catalog, the calibrated cumulative distribution is $C(F(\theta))$. The calibrated posterior quantile at probability level $p$ is therefore $\theta_p=F^{-1}(C^{-1}(p))$, which is the inverse of the $C(F(\theta))$ function.

All credible intervals quoted in Sec.~\ref{sec:results} are calibrated as described above. We validated the procedure on an independent set of 300 mock catalogues, drawn from the O3 posterior samples but not used in the construction of $C(f)$ in eq.(\ref{eq:calibration_map}). As shown in Appendix Fig.~\ref{fig:pp_calibration}, after calibration their coverage curves are consistent with the diagonal. 

\subsection{Stellar-mass BBH population models} \label{sec:mass_model}
This section gives a brief description of the two different population models that describe the primary mass of the binary. The binary component masses $m_1$ and $m_2$, with $m_1>m_2$, are sampled following these models where the mass ratio, $q = m_2/m_1$, is considered to follow a power-law 
\begin{equation}
    p(q) \propto q^\beta,
\end{equation}
similar to the phenomenological models used in the LVK analysis \cite{KAGRA:2021duu}. We assume the mass-model parameters are independent of the redshift. In this analysis, the likelihood, as defined in eq.(\ref{eq:likelihood}), uses the source-frame $\rm M_{cwb}$,
and we do not sample over the redshift distribution 
\begin{equation}
    p(z) \propto \frac{dV_c}{dz} \frac{1}{1+z} (1+z)^\lambda ,
    \label{eq:redshift}
\end{equation}
where $ dV_c/dz$ is the differential comoving volume as determined by cosmology. We use a fixed redshift distribution with $\lambda = 3.3$ corresponding to the low-metallicity star-formation rate \cite{Fishbach:2018edt} while inferring the mass model. This chosen value of $\lambda$ is also consistent with the mean value of redshift distribution reported in the GWTC-4 population analysis \cite{LIGOScientific:2025pvj}. The maximum redshift is set to $z=1.9$ following the distribution of the simulated events. 

\subsubsection{Power Law+Peak (PLP) Mass Model}\label{sec:PLP}
The primary mass distribution follows a power law with a tapering function at low masses. In addition to the power law, the model is described by a Gaussian component and this forms the standard PLP model \cite{Talbot:2018cva} used in the LVK analysis. For detailed explanation see Appendix B1b in \cite{KAGRA:2021duu}.
The \texttt{GWPOPULATION} package is used to implement the PLP model in the analysis \cite{Talbot:2024yqw}. The model parameters which are estimated through \texttt{PopCWB} are listed in Table \ref{tab:model_params}.

\begin{table*}[htb]
	\centering
	   \begin{tabular}{cccc}
			\hline \hline
            \multicolumn{4}{c} {\textbf{Model name: PLP}} \\
            \hline
            Parameter & Prior  & Description & Median value  \\
			\hline
			$\alpha$ & $U(-4,12)$ & Spectral index for the power law of the primary mass distribution &  3.13\\
            $\delta_m$ & $U(0M_{\odot}, 10M_{\odot})$ & Range of mass tapering at the lower end of the mass distribution & 7.1\\
            $\lambda_{peak}$ & $U(0,1)$ & Fraction of BBH systems in the Gaussian component & 0.02\\
            $\sigma_m$ & $U(1M_{\odot}, 10M_{\odot})$ & Width of the Gaussian component in the primary mass distribution & 6.6\\
            $\mu_m$ & $U(20M_{\odot}, 50M_{\odot})$ & Mean of the Gaussian component in the primary mass distribution & 40.4\\
            $\beta$ & $U(-2,7)$ & Spectral index for the power law of the mass ratio distribution & -1.39\\
            \hline \hline
            \multicolumn{4}{c} {\textbf{Model name: cWB-PPI}} \\
            \hline
            $a$ & $U(1,4)$ & Spectral index for low-mass power-law of $M_{co}$ distribution & 2.8\\
            $b$ & $U(1,3)$ & Spectral index for high-mass power-law of $M_{co}$ distribution & 2.72\\
            $M_{break}$ & $U(6M_{\odot}, 20M_{\odot})$ &  Break in the progenitor mass distribution where power law changes & 13.1 \\
            $M_{p}$ & $U(20M_{\odot}, 50M_{\odot})$ &  Progenitor mass at which PPI starts & 40.1 \\
            $\rm M_{BH}^{max} - M_p$ & $U(0.0001 M_{\odot},20 M_{\odot})$ & Maximum black hole mass produced from stellar evolution relative to $\rm M_p$ & 8.1\\
            $\alpha$ & $U(0.0001,0.99)$ & Shape parameter controlling the non-linear relation of $\rm M_{co}$ and $\rm M_{BH}$ & 0.26\\
            $\sigma$ & $U(0.01,1.0)$ & Width of log-normal distribution for $\rm M_{co}$ to $\rm M_{BH}$ mapping & 0.07\\
            $p$ & $U(0,30)$ & Shape parameter controlling the $\rm \overline{PPI}$ contribution & 14.0\\
            $log_{10}(f_{pl})$ & $U[-2.5,0.1]$ & Log of plateau height of $\rm \overline{PPI}$ mass distribution relative to $\rm P_{PPI}(M_{BH}^{max})$  & -1.94\\
            $\beta$ & $U(-2,7)$ & Spectral index for the power law of the mass ratio distribution & 1.6\\
            $\delta_m$ & $U(1 M_{\odot}$, 8$M_{\odot}$) & Range of tapering at lower end of the mass distribution & 5.3 \\
            \hline
		\end{tabular}
		\caption{Table shows the details of the parameters involved in each population model used in this study. The quoted median values are corrected with calibration.}
		\label{tab:model_params}
\end{table*}

\subsubsection{cWB-PPI model}
This section presents a black hole mass model that incorporates the PPI physics along with contributions from dynamically formed black holes. In the literature, main-sequence stars are known to follow a power-law distribution \cite{Salpeter:1955it, Kroupa:2001jy}. However, in this analysis, we will focus on identifying the onset and strength of pulsational pair-instability (PPI), which are primarily determined by the mass and carbon fraction of the progenitor's carbon-oxygen (CO) core~\cite{Woosley:2016hmi, Farmer:2020xne}. Since the exact relationship between the main-sequence mass and the CO core mass of the progenitor is not fully understood, we assume that the CO core mass ($\rm M_{co}$) at the pre-collapse stage follows a broken power-law distribution~\cite{Blanchard:2020wpn, Golomb:2023vxm}
\begin{align}
     P_{\rm co}(\rm M_{\rm co}) \propto
\begin{cases}
     & \rm M_{\rm co}^{-a}, \, \,  \rm M_{\rm co} < M_{\rm break},\\
     & \rm M_{\rm co}^{-b}, \, \,  \rm M_{\rm co} \geq M_{\rm break}
\end{cases}     
\end{align}
where the $\rm M_{break}$ is kept a free parameter. In the following sections, progenitor mass will refer to the carbon-oxygen core mass of the star indicated by $\rm M_{co}$, unless otherwise stated. To derive the black hole mass distribution from the progenitor mass distribution, a mapping between $\rm M_{co}$ and the black hole mass $\rm \bar{M}_{BH}$ is introduced. This mapping relation is motivated by stellar evolution simulation studies, in which the progenitor star is evolved to produce a remnant black hole incorporating PPI phenomenon at different metallicities \cite{Baxter:2021swn, Farmer:2019jed}. 
The mapping of $\rm M_{co}$ to $\rm \bar{M}_{BH}$ assumes the functional form
\begin{equation}
\bar{\rm M}_{\mathrm{BH}}(M_{\mathrm{co}})=
\begin{cases}
\rm M_{\mathrm{co}} & \rm M_{\mathrm{co}}<\rm M_{p}\\[4pt]
\rm M_{\mathrm{co}}+\alpha\Delta\ln\bigl[1-\tfrac{M_{\mathrm{co}}-\rm M_{p}}{\Delta}\bigr]
  & \rm M_{\mathrm{co}}\ge M_{p}
\end{cases}
\label{eq:MI_MBH}
\end{equation}
where, below the mass $\rm M_{p}$, the relation between $\rm M_{co}$ and $\rm \bar{M}_{BH}$ is linear. The parameter $\rm M_{p}$ denotes the onset of the PPI mass regime, beyond which the remnant black hole mass follows a non-linear behavior with $0<\alpha<1$ being the mass function shape parameter and $\Delta$ controlling the width of the PPI region in $\rm M_{co}$ space. The mapping relation holds in the progenitor space till $\rm M_{co} < M_p + \Delta$. The maximum black hole mass obtained from stellar evolution is $\rm M_{BH}^{max} = M_{p} + \Delta (1-\alpha + \alpha\, \ln \alpha)$. The uncertainty arising from unknown physics in the  $\rm \bar{M}_{\mathrm{BH}}(M_{co})$ mass function  is incorporated following the same procedure as in~\cite{Golomb:2023vxm} where the final black hole mass ($\rm M_{BH}$) distribution is defined as
\begin{equation}
 \small
    P_{\mathcal{N}}(\rm M_{\rm BH} | M_{\rm co}) = \mathcal{N} \left[{\rm ln} 
    (\bar{M}_{\rm BH}(M_{\rm co})),\sigma\right]({\rm ln} M_{\rm BH}).
    \label{eq:final_MI_MBH}
\end{equation}
For a fixed progenitor mass $\rm M_{co}$, this distribution incorporates a log-normal scatter with the mean ${\rm ln}(\bar{M}_{BH}$) and standard deviation $\sigma$. The black hole mass distribution governed by PPI is obtained by integrating over the progenitor mass $\rm M_{co}$
\begin{equation}
\small
 P_{\rm PPI}(\rm M_{\rm BH})
  = \int P_{co}(\rm M_{co}) \, P_{\mathcal{N}}(\rm M_{\rm BH} | M_{co}) \,\, dM_{co}.
\label{eq:PPISN}
\end{equation}

The model also takes into account the contribution from the dynamically formed black holes as well as black holes formed from the direct collapse of heavy stars \cite{Bond:1984,Mehta:2021fgz,Sakurai:2017opi}
\begin{equation} 
 \small
  P_{\overline{\rm PPI}}({\rm M}_{\rm BH}) = f_{pl} \, P_{\rm PPI}({\rm M^{max}_{BH}})\, \frac{x^p}{(1+x^p)} \,\, ,  x=\frac{{\rm M_{BH}}}{{\rm M_{BH}^{max}}}.
  \label{eq:no_PPISN}
\end{equation}
The shape parameter `p' controls the behavior of the function which can vary from being a uniform distribution to a step function at $\rm M_{BH}^{max}$. The parameter $f_{pl}$ controls the magnitude of the $\rm \overline{PPI}$ contribution, which equals $f_{pl}/2$ times $\rm P_{PPI}$ at $\rm M_{BH}^{max}$. 
The model in eq.(\ref{eq:no_PPISN}) accounts for black holes formed from both the direct collapse of heavy stars and dynamical formation channel. However, distinguishing between these two populations is not possible with the O3 data since only one event with $\sim 85 M_\odot$ primary mass is observed. For later LVK observing runs, reporting component masses as high as $\sim 137 M_{\odot}$ \cite{LIGOScientific:2025rsn}, it could be possible to explore in more details the high mass black hole spectrum.

Thus, the combined mass distribution is given by 
\begin{equation}
 \small
    p({\rm M_{BH}}) =   \frac{S({\rm M_{BH})}}{\mathcal{Z}} \Big[ P_{\mathrm{PPI}}(\rm M_{BH}) +
                   P_{\overline{\mathrm{PPI}}}({\rm M_{BH}}) \Big] 
    , 
\end{equation}
where the PPI and $\rm \overline{PPI}$ distributions are given in eq.(\ref{eq:PPISN}) and eq.(\ref{eq:no_PPISN}) respectively with $\mathcal{Z}$ being the normalization factor. The function $S(\rm M_{BH})$ denotes the low-mass taper function as used in the PLP model and implemented in this analysis through the \texttt{GWPopulation} package \cite{Talbot:2024yqw}. The model parameters estimated with \texttt{PopCWB} are listed in Table \ref{tab:model_params}. In summary, the parametrization of the cWB-PPI mass model helps extraction of physically meaningful information, in terms of $\rm M_{\rm BH}^{max}$ indicating the lower edge of the black hole mass gap directly from the data.

\begin{figure*}[htb]
    \centering
    \subfloat[]{
  \includegraphics[width=0.47\textwidth]{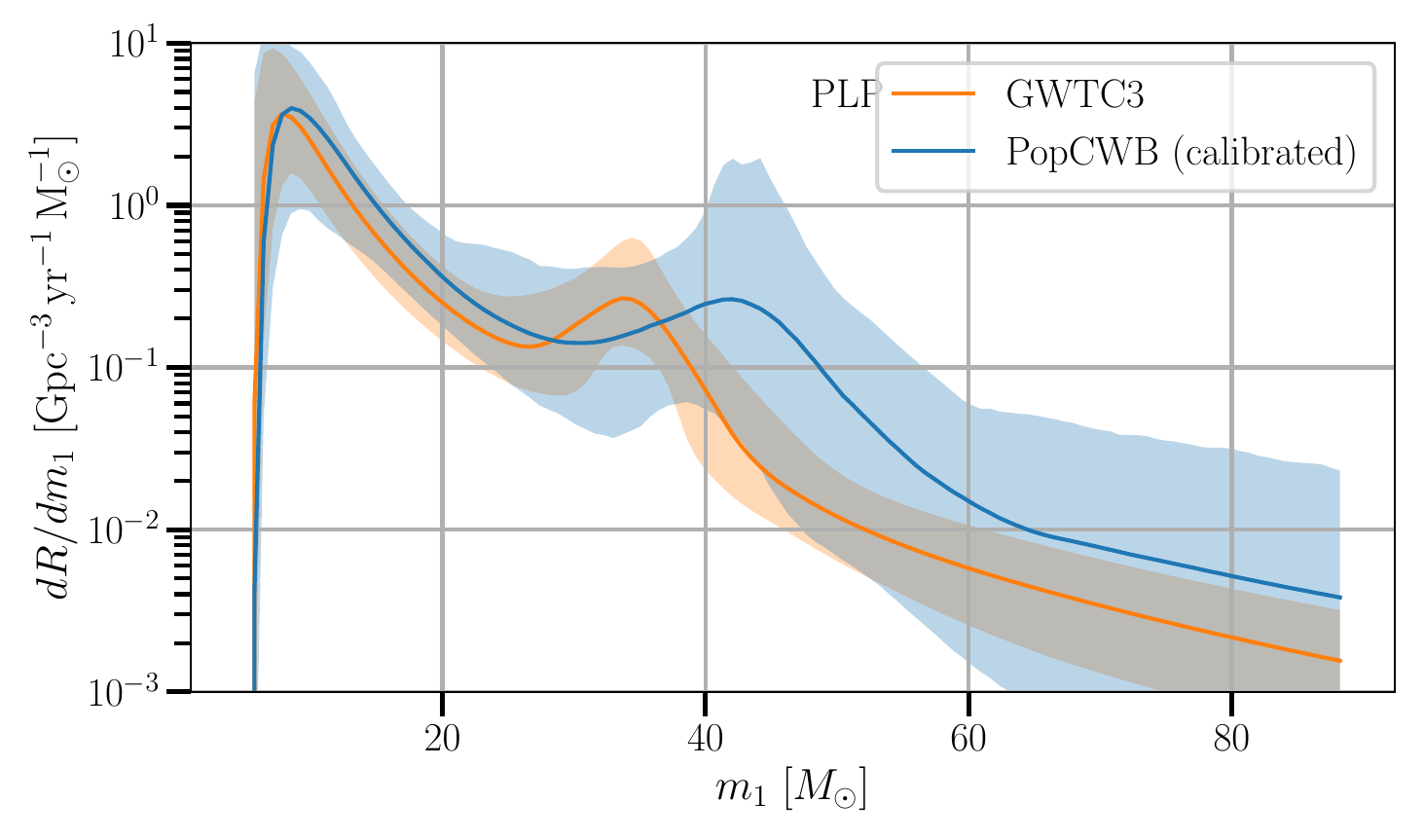}%
    }
   \subfloat[]{
  \includegraphics[width=0.47\textwidth]{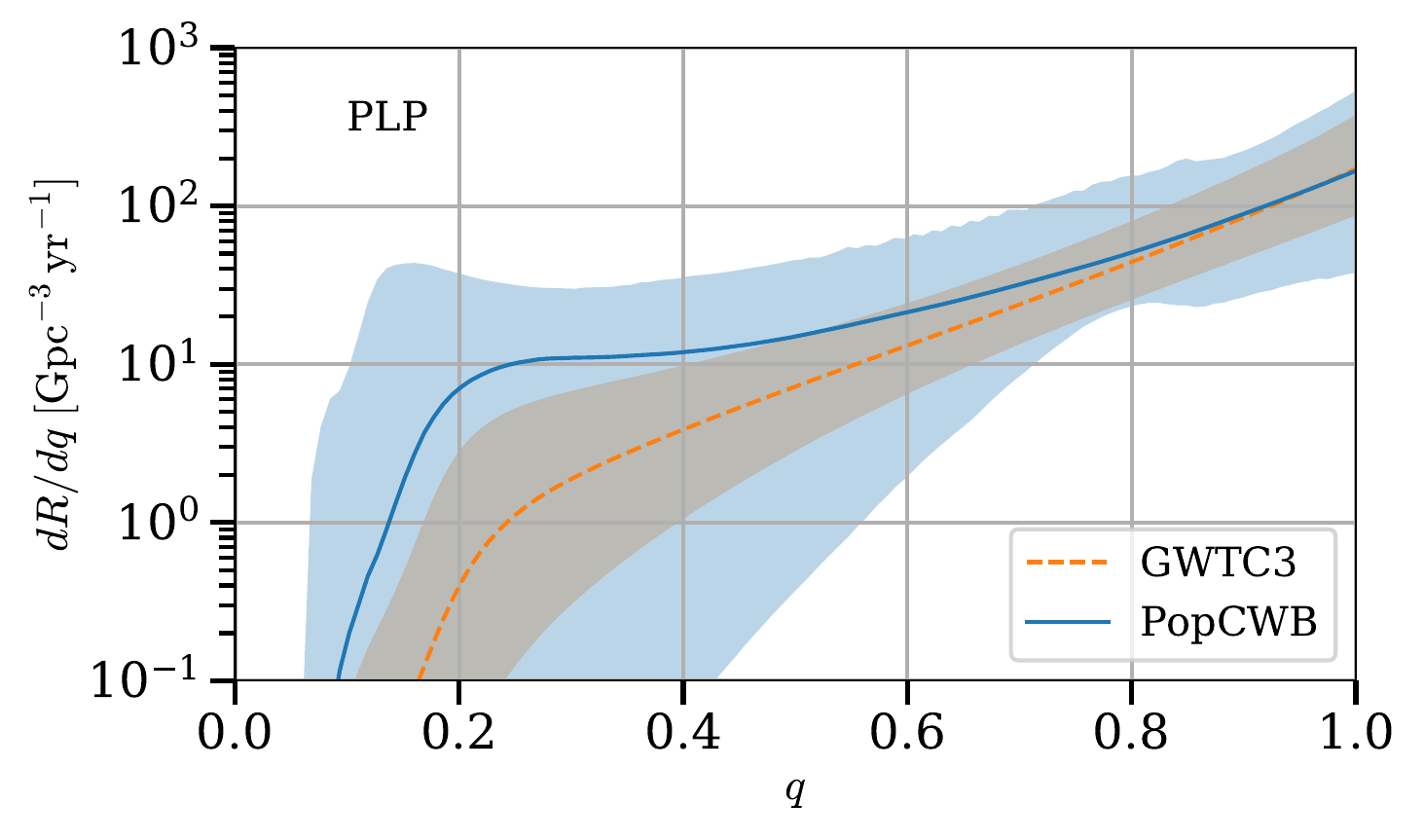}%
  }
    \caption{The plots show the differential merger rate with respect to primary mass and mass ratio as a function of primary component mass of the binary. The figure compares the two analysis frameworks, \texttt{PopCWB} and the standard population inference as reported in GWTC-3 population study \cite{KAGRA:2021duu} for the same population model. The solid curve corresponds to the median values with the shaded region representing $90\%$ credible interval. Within the error bars, both analysis show overall consistency. }
    \label{fig:PLP_rate}
\end{figure*}

\section{Results} \label{sec:results}

In this section, we discuss the inference results for different population models analyzed with the GW events observed during O3 LVK run. The parameters sampled with BILBY for each population model, the priors involved, and the median posterior value after calibration for each model are quoted in Table \ref{tab:model_params}. Note that due to the large parameter space involved, the minimum $5.01 M_{\odot} $ and maximum  $88.22 M_{\odot}$ black hole mass are kept fixed for all models. The fixed values of these parameters are chosen from the results reported for the PLP model in GWTC-3 population analysis \cite{KAGRA:2021duu}.
\subsection{Comparison of population inference frameworks}
We begin by comparing the analysis frameworks, namely the \texttt{PopCWB} algorithm outlined in this study and the standard LVK inference algorithm \cite{KAGRA:2021duu}, using the PLP population model. Fig.\ref{fig:PLP_rate} shows the differential merger rate based on the parameter posteriors obtained from the both analyses. The PLP distribution reported by LVK is renormalized to a physical merger rate using the cWB detection efficiency, calibrated against the 36 O3 events with IFAR $>1$ yr \cite{KAGRA:2021duu, Golomb:2023vxm}. The results indicate overall consistency within the error bars in both the mass and mass ratio distributions. However, the PLP distribution obtained with \texttt{PopCWB} does not observe 
a Gaussian peak with mean $\mu =33.73^{+2.26}_{-3.79} M_{\odot}$ and $\sigma = 3.56^{+4.59}_{-2.10}$, as inferred in the GWTC-3 population analysis where the error bars correspond to $90\%$ credible intervals. Instead, it exhibits 
the Gaussian peak centered around a shifted mean at $ \mu = 40.4^{+7.2}_{-13} M_{\odot}$ with a standard deviation of $\sigma = 6.6^{+3.4}_{-5.3}$ after calibration correction. The mass distribution inferred with \texttt{PopCWB}, which shows an excess around $40 M_{\odot}$, is more consistent with the pileup being created by PPI~\cite{Baxter:2021swn, Farmer:2019jed}.  

On the other hand, the mass ratio distribution is less constrained for the \texttt{PopCWB} analysis. The variations in the mass spectrum and the wider confidence intervals observed in the \texttt{PopCWB} analysis can be attributed to obtaining mass model parameters from a smaller sample of 36 GW events, compared to 69 events analyzed in the GWTC-3 population analysis \cite{KAGRA:2021duu}.

\begin{figure*}[htb]
    \centering
    \subfloat[]{
  \includegraphics[width=0.47\textwidth]{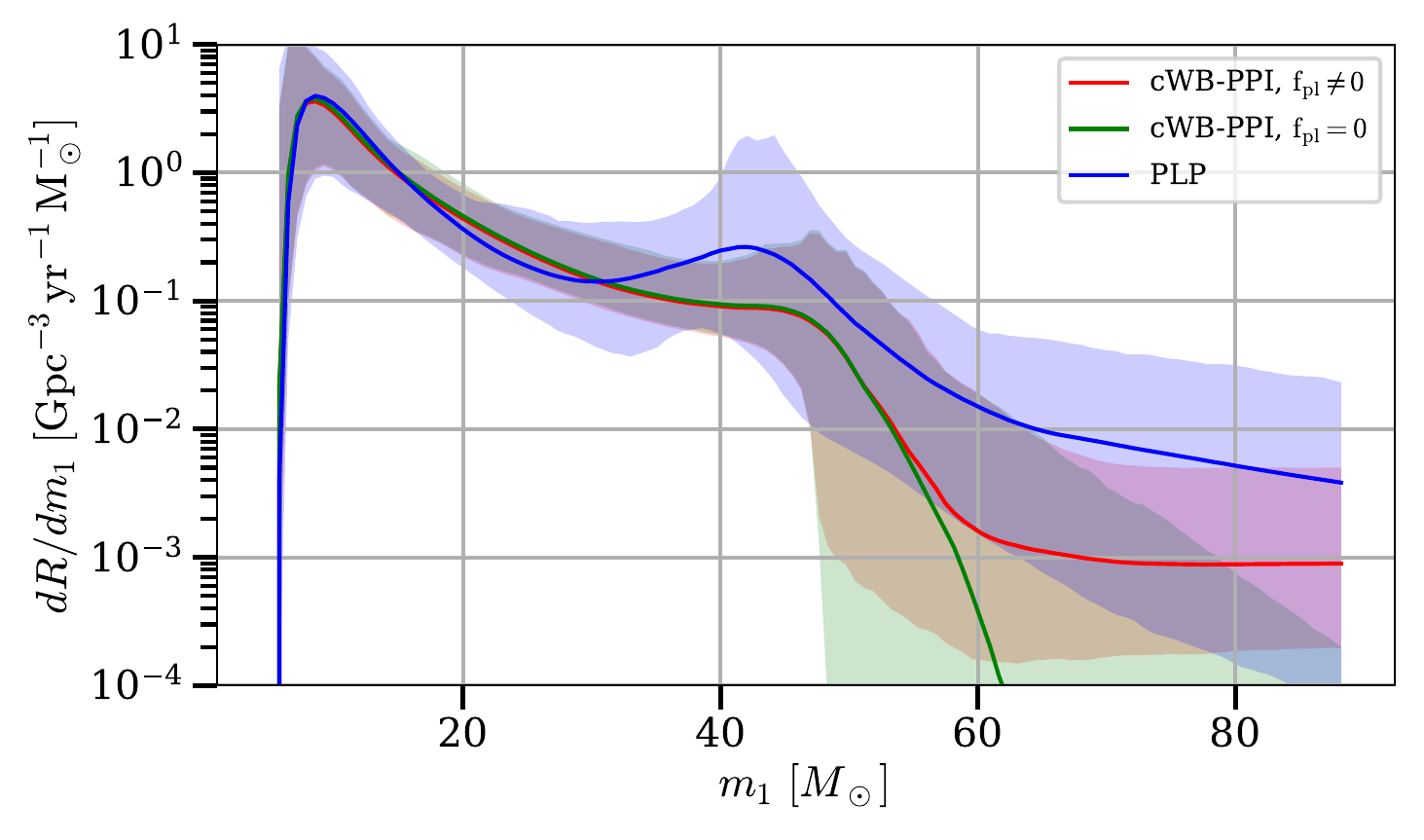}%
    }
   \subfloat[]{
  \includegraphics[width=0.47\textwidth]{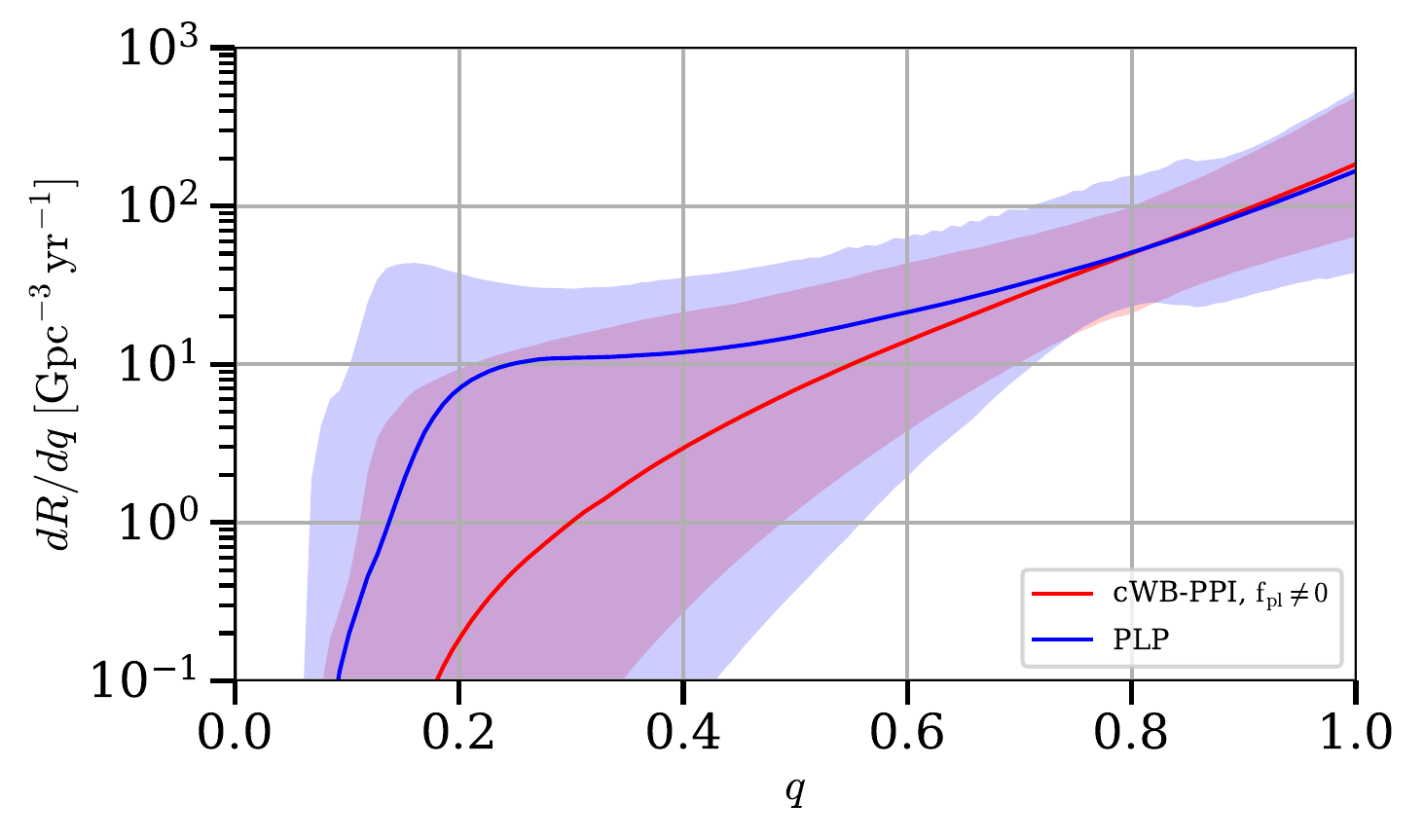}%
  }
    \caption{The plots show the differential merger rate with respect to primary mass and mass ratio as a function of primary component mass of the binary for the models analysed in this work. The population parameters are taken from the analysis done with \texttt{PopCWB}. The solid curve corresponds to the median values with the shaded region representing $90\%$ credible interval. Both models show overall consistency within error bars even though each model has slight differences in the features present in mass distribution (see Sec.\ref{sec:PopCWB_compare} for a detailed discussion). Note that the green curve in the primary mass distribution represents the differential rate for cWB-PPI model when the $\rm \overline{PPI}$ contribution is turned off by setting $f_{pl} =0$. }
    \label{fig:PopCWB_compare}
\end{figure*}

\subsection{Results from cWB-PPI model } \label{sec:PopCWB_compare}
The cWB-PPI model is parametrized by 11 parameters describing the primary mass and mass ratio distributions as shown in Table \ref{tab:model_params}. Their values are quoted in this section with calibration and error bars representing $90\%$ credible intervals. The progenitor distribution at low masses is defined by a power law with $a = 2.8^{+1.1}_{-1.6}$, which remains prior-dominated, as the O3 data contain only a few low-mass events detected by cWB.  Beyond ${\rm M_{break}} = 13.1^{+6.5}_{-6.5} M_{\odot}$, which also has a weakly constrained posterior, the progenitor masses follow a distribution with a power law index of \( b = 2.72^{+0.28}_{-0.98} \). 

The PPI phenomenon begins to dominate the progenitor mass spectrum from \( {\rm M_{p}} = 40.1^{+9.7}_{-12.5} M_{\odot} \) and extends to \( {\rm M_{p}} + \Delta = 60^{+30}_{-14} M_{\odot}\), 
the CO core mass at which the PISN occurs.  In the literature, the accepted limit for the onset of the PISN regime is \( {\rm M_{co}} \geq 64 M_{\odot} \) \cite{Wright:2017zyq}. Nevertheless, our analysis does not rule out progenitors with masses significantly larger than \( 64 M_{\odot} \) from contributing to black hole formation. We anticipate that constraints on \( {\rm M_{p}} + \Delta\) will improve as more GW events from later LVK observing runs are included. 

Our analysis indicates that all progenitors with masses between \( \rm M_{p} \) and \( \rm M_{p} + \Delta \) contribute to a narrow black hole mass range determined by $\rm M_p$ and $\rm M_{BH}^{max}$. 
The maximum black hole mass formed through stellar evolution is \( {\rm M_{BH}^{max}} = 48.3^{+8.2}_{-10.2} M_{\odot} \). This value agrees with the predicted maximum black hole mass derived from the stellar evolution simulations that take into account the \( \rm ^{12}C(\alpha, \gamma) ^{16}O \) reaction rates \cite{Baxter:2021swn, Farmer:2019jed}. Following the relation between the \( \rm ^{12}C(\alpha, \gamma) ^{16}O \) rate and the maximum black hole mass in \cite{Mehta:2021fgz}, and our estimated \( {\rm M_{BH}^{max}} \), we conclude that the \( \rm ^{12}C(\alpha, \gamma) ^{16}O \) reaction rate cannot be significantly lower than the nominal reaction rate.

The inference of the model parameters describing dynamical formation and direct collapse remain unconstrained, as GW190521 \cite{LIGOScientific:2020iuh, LIGOScientific:2020ufj} is the only high-mass event recorded during the O3 observing run. The parameters \( p \) and \( f_{pl} \) have weakly constrained posteriors, resulting in a broad error region around the high-mass tail, as illustrated in Fig.~\ref{fig:PopCWB_compare}. The mass ratio distribution, modeled as a power-law, remains consistent with the PLP model when analyzed using \texttt{PopCWB} inference framework. 

\begin{figure}[hbt]
    \centering
  \includegraphics[width=0.47\textwidth]{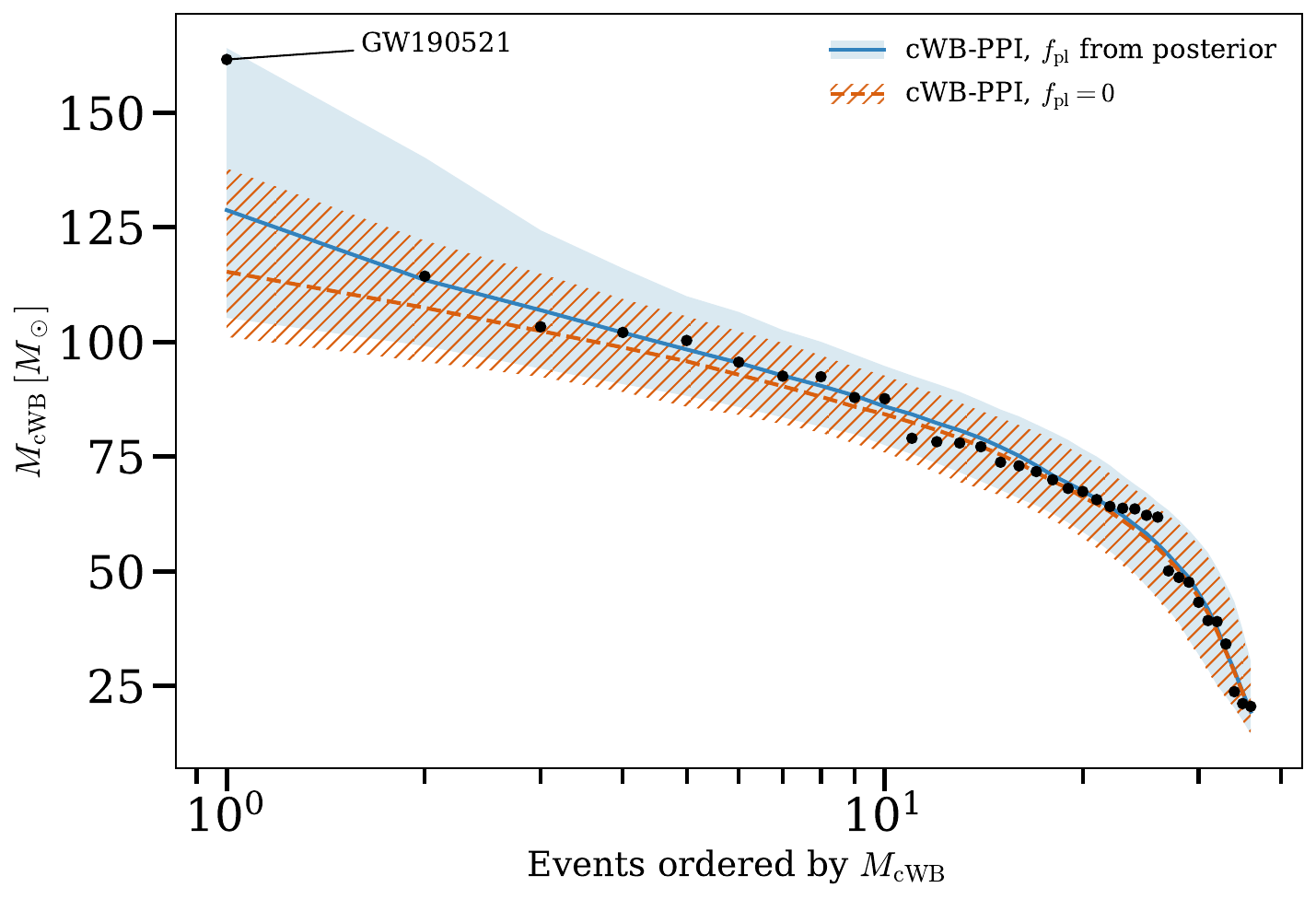}%
    \caption{The figure shows the consistency of the observed events (black dots) with the cWB-PPI model (blue band) which includes the contribution of the dynamically formed black holes (see eq.(\ref{eq:no_PPISN})). The model parameters correspond to their calibrated median values. However, when the dynamical contribution is turned off by setting $f_{pl} = 0$, the red-dashed plot shows GW190521 being an outlier in comparison to stellar evolution. This indicates possible support for the GW190521 being formed dynamically. }
    \label{fig:outlier}
\end{figure}

The presented population analysis also provides insight into a possible formation scenario for the primary component of GW190521. As shown in Fig.~\ref{fig:outlier}, GW190521 is well-described by the model when it includes the contribution of $\rm \overline{PPI}$ as shown in eq.(\ref{eq:no_PPISN}). However, when this contribution is excluded by setting \( f_{pl} = 0 \), GW190521 falls far outside the confidence interval of the model, making it a clear outlier. Multiple studies suggest that the primary component of GW190521 is unlikely to have formed through stellar evolution \cite{OBrien:2021sua, LIGOScientific:2020iuh}, a conclusion our analysis also supports.

All events other than GW190521 are well described by 
the $\rm {PPI}$ part of the primary mass spectrum shown in Fig.~\ref{fig:PopCWB_compare}, which falls off rapidly above $48 M_{\odot}$. The dispersion of the \( {\rm M_{BH}^{max}} \) mass cutoff is described by the model parameter $\sigma=0.07^{+0.17}_{-0.06}$ which accounts for possible differences in the progenitor conditions, including metallicity and stellar wind mass loss \cite{Farmer:2019jed,Spera:2017fyx,Vink:2020nak,Belczynski:2019fed, Winch:2024xdt}, rotation \cite{Marchant:2020haw,Mapelli:2019ipt} and the stellar binary interaction history including accretion~\cite{Marchant:2018kun,vanSon:2020zbk}.  Because each progenitor can have a distinct evolution pathway, the resulting black hole masses may deviate from the \( {\rm M_{BH}^{max}} \) 
representing the characteristic low mass limit of the PISN mass gap.

The local merger rate at $z=0$ is calculated for each model analysed with  \texttt{PopCWB} assuming an underlying redshift distribution following a power-law with spectral index 3.3. For PLP, the local merger rate is $30.4^{+47.3}_{-16.6}$ $\rm Gpc^{-3} yr^{-1}$ while it is $29.1^{+32.8}_{-16}$ $\rm Gpc^{-3} yr^{-1}$ for cWB-PPI.  We observe the rates to be consistent within the error bars for both models and also with the GWTC-3 BBH rate \cite{KAGRA:2021duu}. Since the local rate is dominated by the well-measured mid-mass population, this agreement indicates that the astrophysically parameterized mass model does not bias the rate estimates.

In order to quantify the difference between the models and to ascertain the most preferred model, the Bayes factor is calculated. The natural logarithm of the Bayes factor is $\rm ln \mathcal{B}^{PLP}_{cWB-PPI} = -2.72$ which is showing a moderate preference for the cWB-PPI model. 

\section{Conclusion and discussion} \label{sec:conclusions}

In this work, we have presented \texttt{PopCWB}, a Bayesian framework
designed to infer the parameters of the BBH population mass distribution. The standard population inference used in literature takes an inverse approach by using the posterior samples obtained from parameter estimation of individual events to infer the underlying population model. On the other hand, \texttt{PopCWB} takes a forward approach by varying the population models and using a large simulation set to infer the most optimal model parameters best describing the detected events. The simulated BBH events and observed GW events in O3 LVK run, which are reconstructed by the cWB pipeline with IFAR$>1$ yr are used in the analysis. The total mass of all events is estimated with a machine-learning regression and used in the construction of the likelihood as shown in eq.(\ref{eq:likelihood}).

The \texttt{PopCWB} algorithm is used to make astrophysically relevant observations about the black hole mass spectrum.
First, we compared the differential merger rate distribution for the PLP model when analysed with \texttt{PopCWB} to the GWTC-3 population results \cite{KAGRA:2021duu}. We observed an overall consistency within error bars. However, the \texttt{PopCWB} PLP inference shows the excess near $\sim40\,M_\odot$ and not around $\sim35\,M_\odot$. The peak shifted at $\sim40\,M_\odot$ is more consistent with the pileup being created by the PPI phenomenon. Second, we inferred the parameters of an astrophysically motivated model, cWB-PPI, with \texttt{PopCWB}. The model takes into account the PPI phenomenon while describing the formation of black hole from the carbon-oxygen progenitor with mass $\rm M_{co}$. We observe that the start of the PPI regime in $\rm M_{co}$ space is at $\sim 40 M_{\odot}$ while the end of the PPI region remains unconstrained. Our analysis does not rule out progenitors with ${\rm M_{co}} > 64 M_{\odot}$ contributing to black hole formation. The inferred maximum black hole mass formed from stellar evolution in the cWB-PPI model is $\sim48\,M_\odot$. This value
is compatible with the predicted maximum black hole mass derived from the stellar evolution simulations taking into account the $\rm ^{12}C(\alpha, \gamma) ^{16}O$ reaction rates \cite{Baxter:2021swn, Farmer:2019jed}. Within our analysis
the excess at $\sim48\,M_\odot$ can be associated with the PPI pileup
without invoking anomalous reaction rates. An estimation of the Bayes factor showed a moderate preference for the cWB-PPI model over the PLP model.

Our next step is to apply \texttt{PopCWB} to the substantially
larger catalogue of BBH events accumulated during O4 run, where additional high-mass
detections should sharpen the constraints on the cWB-PPI model parameters and begin to
inform the high-mass subpopulation. The \texttt{PopCWB} can be further extended to include the evolution of the model parameters with redshift, which is left for future work.

\section*{Acknowledgment}
This material is based upon work supported by NSF's LIGO Laboratory, which is a major facility fully funded by the National Science Foundation. We gratefully acknowledge the support of LIGO and Virgo for the provision of computational resources, especially LIGO Laboratory, which is supported by the NSF Grant No. PHY $0757058$ and PHY $0823459$. This work was supported by the NSF grant No. PHY-2409372. This research has made use of data, software, and/or web tools obtained from the Gravitational Wave Open Science Center, a service of LIGO Laboratory, the LIGO Scientific Collaboration, and the Virgo Collaboration. We acknowledge the valuable feedback from Marek~J.~Szczepa\'nczyk, Gayathri.V and Matthew Mould. For the purpose of open access, the authors have applied a Creative Commons Attribution (CC BY) licence to any Author Accepted Manuscript version arising. This document has LIGO preprint number P2600508.

\bibliographystyle{apsrev}
\bibliography{ref}	

\appendix
\begin{figure}[t]
    \centering
    \includegraphics[width=0.88\linewidth]{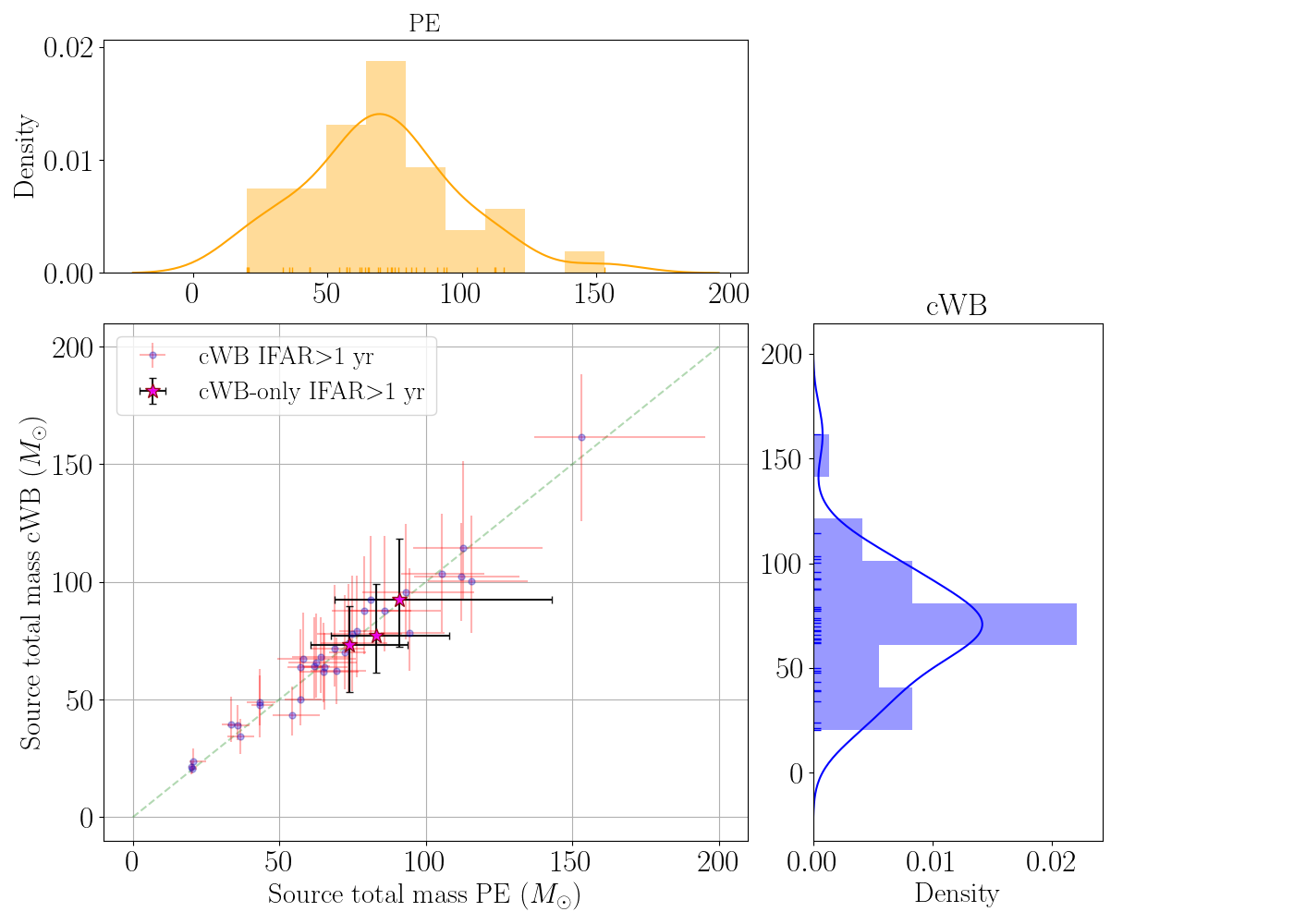}
    \caption{Figure shows the comparison of the cWB source-frame total mass point estimate with respect to the results reported from parameter estimation. The 90\% confidence intervals for both methods are denoted by solid lines around the marker for each event. The cWB-only events detected in O3 are highlighted with star markers and black lines for showing error bars. The density distributions for the cWB and PE estimates are plotted in diagonal plots.}
    \label{fig:xgb_mass_compare}
\end{figure}
\section{XGBoost training and testing}\label{sec:xgb_training}
We utilize the XGBoost\cite{Chen_2016} regression algorithm to estimate the source-frame total mass $M_{cwb}$ of detected binary merger signals. To train the regression model, we follow a similar setup as shown in \cite{OBrien:2021sua} with updates to the set of cWB summary statistics used as input features, injection data considered for training~\ref{sec:injections}, and updated model hyper-parameters. The injection set is divided into two equal training and testing datasets, the testing dataset is used to evaluate the accuracy of the  trained model in estimating $M_{cwb}$. Since XGBoost provides a point estimate of $M_{cwb}$, we use the injections recovered by cWB with IFAR $> 1$ yr from the testing dataset to quantify the 90\% confidence interval. A detailed description of the XGBoost method updates and error estimation on $M_{cwb}$ can be found in \cite{Mishra_Thesis}.

To validate the XGB model performance, we compare the cWB-reconstructed source-frame total mass ($M_{cwb}$) point estimate with the corresponding source-frame total mass provided by conventional parameter estimation $M_{PE}$ (PE) as reported in the GW catalog ~\citep{KAGRA:2021vkt}, plotted in Fig.\ref{fig:xgb_mass_compare}. We find that the cWB estimation for source-frame total mass is in agreement with the reported PE values for all the 36 cWB detections in O3~\citep{Mishra:2024zzs}.

\section{Results from \texttt{PopCWB}}
\begin{figure*}[tb]
    \centering
    \includegraphics[width=0.9\linewidth]{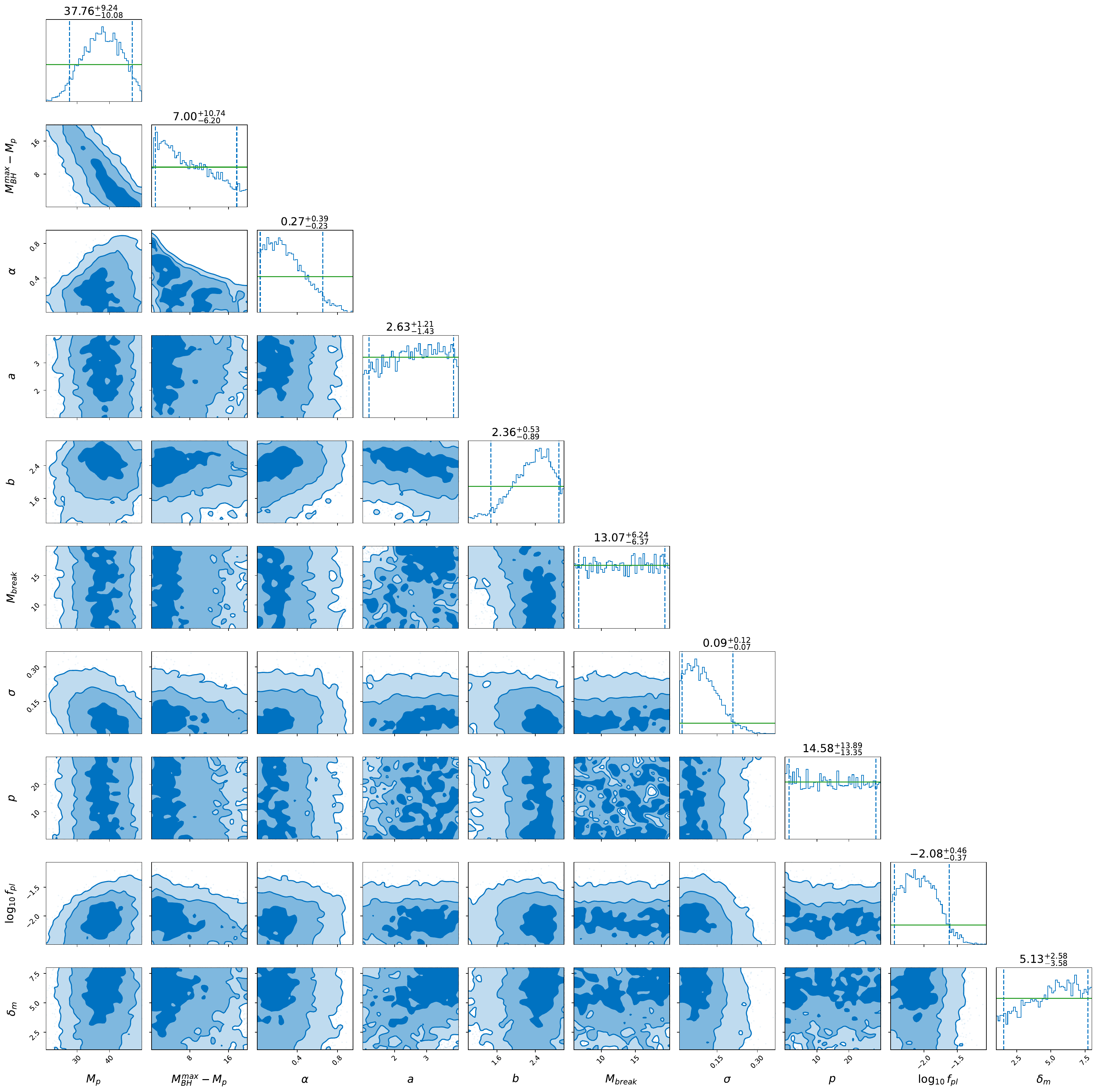}
    \caption{The figure shows the posteriors for the cWB-PPI model parameters inferred with \texttt{PopCWB} without calibration correction. The error bars correspond to 90\% credible interval. The green line shows the uniform prior for all the parameters.}
    \label{fig:cWB_PPISN_corner}
\end{figure*}

In Fig.\ref{fig:cWB_PPISN_corner}, we report the full corner plot for all the sampled parameters corresponding to the cWB-PPI model analysed in this work. The values of the parameters quoted are not calibrated and the error bars correspond to $90\%$ credible intervals.

\section{Calibration Plot}

\begin{figure*}[htb]
    \centering
    \subfloat[]{
  \includegraphics[width=0.47\textwidth]{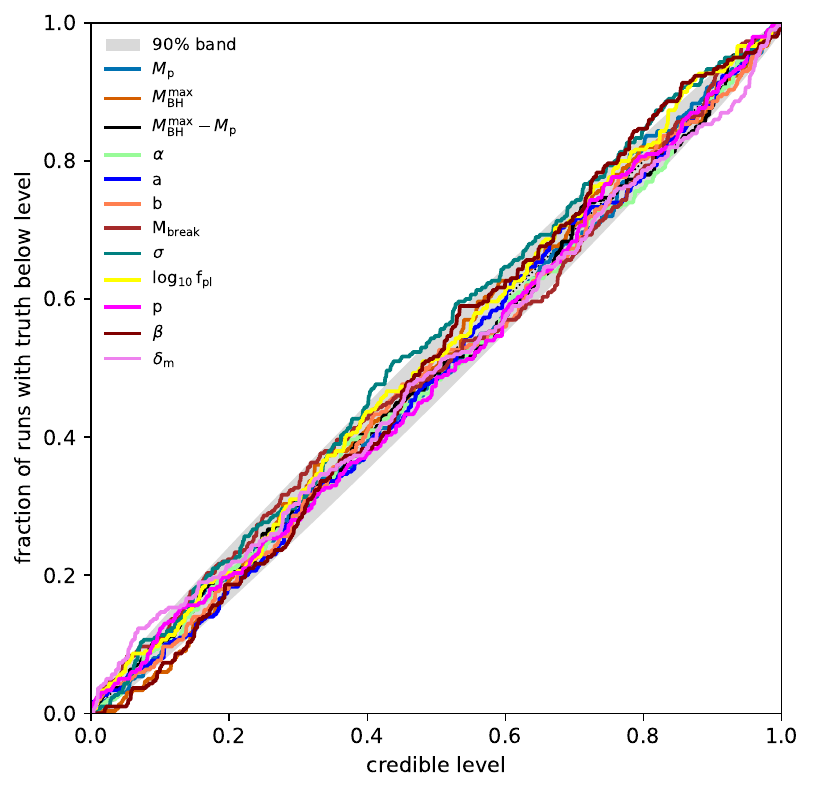}%
    }
   \subfloat[]{
  \includegraphics[width=0.47\textwidth]{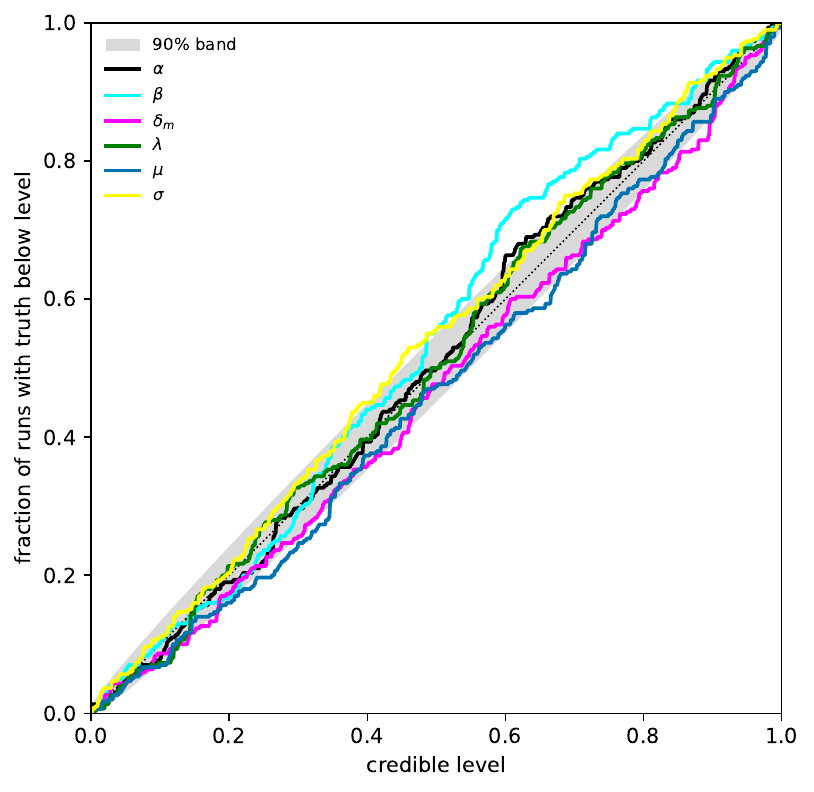}%
  }
    \caption{The curves show the p-p plot for parameters from the cWB-PPI mass model (on left) and PLP model (on right) after calibration. All parameters lie within the 90\% grey shaded region corresponding to 300 runs where true parameter values are drawn from the posterior. }
    \label{fig:pp_calibration}
\end{figure*}

\end{document}